\documentclass[twocolumn,superscriptaddress,showpacs,preprintnumbers,nofootinbib,amsmath,amssymb,floatfix,aps]{revtex4-1}
\usepackage{graphicx}
\usepackage{isotope}
\usepackage{subfigure}
\usepackage{pstricks}
\usepackage{color}

\newcommand{\n}[1]{\ensuremath{|\mathbf{#1}|}}

\graphicspath{{plot/}}

\begin{document}

\title{Estimate of the theoretical uncertainty of the cross sections \\for nucleon knockout in neutral-current neutrino-oxygen interactions}
\author{Artur M. Ankowski}
\affiliation{Department of Physics, Virginia Tech, Blacksburg, Virginia 24061, USA}
\author{Maria B. Barbaro}
\affiliation{Dipartimento di Fisica, Universit\`a di Torino\\
and INFN, Sezione di Torino, Via P. Giuria 1, I-10125 Torino, Italy}
\author{Omar Benhar}
\affiliation{INFN and Department of Physics,``Sapienza'' Universit\`a di Roma, I-00185 Roma, Italy}
\affiliation{Department of Physics, Virginia Tech, Blacksburg, Virginia 24061, USA}
\author{Juan A. Caballero}
\affiliation{Departamento de F\'{\i}sica At\'{o}mica, Molecular y Nuclear, Universidad de Sevilla, 41080 Sevilla, Spain}
\author{Carlotta~Giusti}
\affiliation{Dipartimento di Fisica, Universit\`a degli Studi di Pavia, and INFN, Sezione di Pavia, Via A. Bassi 6, I-27100 Pavia, Italy}
\author{Ra\'ul Gonz\'alez-Jim\'enez}
\affiliation{Departamento de F\'{\i}sica At\'{o}mica, Molecular y Nuclear, Universidad de Sevilla, 41080 Sevilla, Spain}
\affiliation{Department of Physics and Astronomy, Ghent University, Proeftuinstraat 86, B-9000 Gent, Belgium}
\author{Guillermo D. Megias}
\affiliation{Departamento de F\'{\i}sica At\'{o}mica, Molecular y Nuclear, Universidad de Sevilla, 41080 Sevilla, Spain}
\author{Andrea Meucci}
\affiliation{Dipartimento di Fisica, Universit\`a degli Studi di Pavia, and INFN, Sezione di Pavia, Via A. Bassi 6, I-27100 Pavia, Italy}

\date{\today}%

\begin{abstract}
Free nucleons propagating in water are known to produce $\gamma$ rays, which form a~background to the searches for diffuse supernova neutrinos and sterile neutrinos carried out with Cherenkov detectors. As a consequence, the process of nucleon knockout induced by neutral-current quasielastic interactions of atmospheric (anti)neutrinos with oxygen needs to be under control at the quantitative level in the background simulations of the ongoing and future experiments. In this paper, we provide a quantitative assessment of the uncertainty associated with the theoretical description of the nuclear cross sections, estimating it from the discrepancies between the predictions of different models.

\end{abstract}

\pacs{13.15.+g, 25.30.Pt}
%


\maketitle

\section{Introduction}

Large water-Cherenkov detectors have proven to be a powerful tool to address a number of outstanding physics issues. Ongoing and future research programs at Super- and Hyper-Kamiokande~\cite{ref:Hyper-K} include searches of diffuse supernova neutrinos (DSN)~\cite{ref:Super-K_DSN_new,ref:Super-K_DSN_nTagging} and sterile neutrinos in the beam of the T2K experiment~\cite{ref:Ueno}.

These studies require that the backgrounds \cite{ref:SWLi} related to atmospheric-neutrino interactions be estimated with a challenging level of accuracy, because the signals are expected to be elusive---a~few DSN events per year in the fiducial volume of the Super-Kamiokande (SK) detector.

The DSN flux consists of neutrinos and antineutrinos of all the flavors, produced in the past core-collapse supernova explosions. As in the scale of the whole Universe such events take place approximately every second~\cite{ref:Burrows_nature} and have isotropic distribution, the DSN signal is believed to be steady in time and uniform in space~\cite{ref:Lunardini,ref:Lunardini&Tamborra}. Its measurement would shed light on average supernova features impossible to obtain otherwise, the core-collapse event rate within our Galaxy being only a~few per century~\cite{ref:Diehl_SN_rate,ref:Adams_SN_rate}. Therefore, the DSN signal provides a unique window on the bulk picture of the entire supernova population, the understanding of which is essential to address open questions of paramount importance, such as the origin of heavy chemical elements and the production of cosmic rays~\cite{ref:Beacom_AnnuRev}.

The DSN search of the SK experiment is currently performed in the energy range $13\lesssim E_\nu\lesssim90$ MeV~\cite{ref:Super-K_DSN_new,ref:Super-K_DSN_nTagging}, in which the dominant interaction mechanism, sensitive to $\bar\nu_e$'s only, is inverse $\beta$ decay of free protons in the water molecule~\cite{ref:Fogli},
\begin{equation}
\label{eq:DSN_signal}
\bar\nu_e+p\rightarrow e^++n.
\end{equation}

Until recently~\cite{ref:Super-K_DSN_new}, neutrons could not be observed and the only signature of the DSN event~\eqref{eq:DSN_signal} was the signal of the positron, which in water-Cherenkov detectors cannot be distinguished from the signal of an electron or a $\gamma$ ray. As a consequence, all processes yielding $e^+$'s, $e^-$'s, or $\gamma$'s in the relevant energy range, generated backgrounds to the DSN search.

At present~\cite{ref:Super-K_DSN_nTagging}, the upgraded electronics of the SK detector allows neutrons to be detected by the measurement of 2.2 MeV $\gamma$ rays,
resulting from their capture on free protons, with efficiency $\sim$18\%. The requirement that the positron-like signal be followed by the neutron-capture signature has led to a~sizable background reduction, enabling the DSN search to be performed down to energy $\sim$13~MeV~\cite{ref:Super-K_DSN_nTagging}, compared to $\sim$17~MeV of the previous analysis~\cite{ref:Super-K_DSN_new}.

Further progress will be possible if the gadolinium doping program, which may boost the neutron capture efficiency to $\sim$90\%, will be successfully completed in the SK experiment~\cite{ref:Beacom_Ga,ref:Vagins_Ga,ref:Super-K_Ga}. Within this scenario, the neutron produced in reaction~\eqref{eq:DSN_signal} will be signaled by the 8 MeV $\gamma$-ray cascade from the capture in the gadolinium nucleus, easier to observe than the 2.2 MeV emission from the absorption on free protons. Efficient neutron detection will dramatically reduce many backgrounds~\cite{ref:Super-K_DSN_new}, allowing the DSN search to be  extended further into the low-energy region, of crucial importance. However, this is not the case for the contribution of mechanisms producing $\gamma$ rays in coincidence with the neutrons, for which precise estimates of the cross sections and their uncertainties are required.

In this context, the most significant role is played by neutron knockout from oxygen induced by
neutral-current (NC) quasielastic (QE) scattering of atmospheric neutrinos and antineutrinos~\cite{ref:Bays_thesis}. This reaction, occurring at a~rate of $\sim$2 events per day in the fiducial volume of Super-Kamiokande~\cite{ref:Beacom_Ga}, may yield $\gamma$ rays through two mechanisms: (i) deexcitation of the residual nucleus and (ii) interactions of the knocked-out nucleon with the surrounding medium.
The associated cross sections and their uncertainties have been estimated in Refs.~\cite{ref:gamma} and~\cite{ref:nKnockout} within the approach based on the impulse approximation scheme and realistic nuclear spectral functions~\cite{ref:Omar_LDA,ref:Omar_oxygen}. In the following, we will refer to it as the SF approach. It is worth noting that the predictions of Refs.~\cite{ref:gamma,ref:nKnockout} have been found consistent with the experimental result of the T2K Collaboration~\cite{ref:T2K_NC}.

To gauge the uncertainties associated with the theoretical description of the nucleon-knockout cross sections, in this paper we compare the results of the SF formalism~\cite{ref:Omar_LDA,ref:Omar_oxygen} to those obtained from different approaches, namely relativistic plane-wave impulse approximation (RPWIA), relativistic mean-field (RMF)~\cite{Maieron:2003df,Caballero:2005sn}, relativistic Green's functions (RGF)~\cite{Capuzzi:1991qd,Meucci:2003cv}, and superscaling (SuSA)~\cite{Amaro_SuSA_CC,Amaro_SuSA_NC}. For the sake of completeness, we also show the results of the relativistic Fermi gas (RFG) model of Ref.~\cite{Amaro_SuSA_NC}.

Motivated by the requirements of  both the ongoing search of the DSN signal in SK and the determination of the total active-neutrino flux in the T2K experiment,
we consider both neutron and proton knockout from oxygen induced by (anti)neutrino NC QE interaction.
Note that NC interactions are not sensitive to neutrino flavor. Hence, oscillations between active neutrinos do not affect the rate of NC events detected with a~near and a far detector. Should a deficit of NC events in the far detector be observed, this would point to oscillations into sterile neutrinos. To determine the rate of NC QE interactions in the kinematical setup of T2K, it is necessary to measure the associated $\gamma$-ray production, owing to the $\sim$50\% contribution of neutron knockout and high Cherenkov threshold of protons in water~\cite{ref:Ueno}.

The different theoretical approaches employed to describe neutrino- and antineutrino-nucleus interactions are outlined in Sec.~\ref{approaches}, while Sec.~\ref{results} is devoted to the
discussion of the calculated NC QE cross sections for oxygen. Finally,  in Sec.~\ref{conclusions}, we summarize our findings and state the conclusions.

\section{Theoretical approaches}
\label{approaches}

In scattering processes off nuclei, the internal structure of the target is probed with space resolution $\sim{2\pi}/\n q$, where $\n q$ denotes the momentum transfer.
Therefore, for $\n q$ larger than  $\sim{2\pi}/d$, $d$ being the average interparticle distance in the target, nuclear scattering is expected to reduce to the incoherent sum of elementary
processes, involving bound moving nucleons~\cite{ref:Omar_FSI_NM}. This is the premise underlying the impulse approximation (IA), providing the conceptual
framework of all approaches discussed in this work.

\subsection{Superscaling}

The SuperScaling Approch (SuSA) to neutrino scattering, proposed in Refs. \cite{Amaro_SuSA_CC,Amaro_SuSA_NC} for CC and NC reactions, respectively,
exploits the measured electron-nucleus cross sections to predict neutrino-nucleus cross sections.

The analysis of electron scattering data in the QE sector \cite{DS1,DS2} shows that, at large momentum transfer,
the reduced inclusive cross section (i.e. the nuclear cross section divided by the sum of the elementary electron-nucleon cross sections) is largely independent of both $|{\bf q}|$  (scaling of the first kind) and the nuclear target (scaling of the second kind), when
represented as a function of the scaling variable $\psi$.

These properties are clearly observed in the longitudinal channel, whereas violations associated with reaction mechanisms not taken into account within the IA scheme---such as inelastic scattering and processes involving meson-exchange currents---show up in the transverse channel, mainly at energy transfer larger than that corresponding to
single-nucleon knockout.

On the basis of the above observations, a phenomenological superscaling function $f(\psi)$ has been extracted from electron-scattering data within a fully relativistic framework.
The function $f(\psi)$ embodies the essential nuclear dynamics, including both initial- and final-state physics. Its most striking features are a pronounced asymmetric tail at large
energy transfer and a maximum $\sim$20\% lower than the RFG prediction.

Within SuSA, neutrino-nucleus cross sections are simply obtained multiplying the function $f(\psi)$ by the appropriate elementary weak cross sections.
Although phenomenological, this approach has several merits: (i) it agrees by construction (up to scaling violations) with electron scattering data,
(ii) it accounts for both kinematical and dynamical relativistic effects---which are known to be significant even at moderate momentum and energy
transfer---and can therefore be safely used in a broad energy range, and (iii) owing to scaling of second-kind, it allows for a consistent treatment of different
target nuclei. In addition, it is worth mentioning that the model has been successfully  extended to higher energies, well beyond the QE regime \cite{nonQE}.

\subsection{Relativistic Approaches}

The RPWIA, RMF and RGF approaches provide a microscopic and fully relativistic description of the scattering process, at both
kinematical and dynamical level.

In all cases,  the bound nucleon states are represented by four-spinors, obtained from the self-consistent
solution of the Dirac-Hartree equation  derived from a Lagrangian including $\sigma$, $\omega$ and $\rho$ mesons within the  mean-field
approximation  \cite{sigma-omega_1,sigma-omega_2}.
The state of the outgoing nucleon is described by a
relativistic (four-spinor) scattering wave function.

The RPWIA approach can be regarded as the simplest implementation of the above formalism. Within RPWIA all final-state
interactions (FSI) between the nucleon interacting with the beam particle
and the spectators are neglected, and the knocked out particle is described by a plane wave. This approximation, while leading to a
dramatic simplification  in the description of the process,  results in the appearance of significant discrepancies between the calculated cross sections and
scaling functions and the data. For example, the RPWIA scaling
function does not exhibit the asymmetry clearly visible in the data, thus suggesting that
a more realistic description, taking into account the effects of FSI, is needed.

In this work, we have used two different schemes, referred to as RMF and RGF. The former makes use of relativistic distorted waves obtained
with the same relativistic scalar and vector energy-independent potentials used to determine the initial bound states.
The results of this approach are in excellent agreement with the phenomenological scaling functions extracted from
electron-nucleus scattering data \cite{Maieron:2003df,scalcab}.  Moreover, the transverse scaling function exhibits an enhancement  of
$\sim$20\%, with respect to the longitudinal one \cite{LT}. It has been suggested that a similar enhancement may have a significant
impact on the measured neutrino cross sections \cite{TEN1,TEN2,TEN3,TEN4}.

In the RGF approach, FSI are included using the relativistic Green's function formalism and a
complex optical potential \cite{RGF1,RGF2,RGF3}.
The use of a complex optical potential allows for a fully consistent description of the inclusive response, as the loss of flux associated with
the occurrence of inelastic processes is taken into account. Moreover, because of the analyticity properties of the optical potential, the Coulomb sum rule
is fulfilled by construction \cite{RGF1,RGF4}.

Like the RMF approach, the RGF provides a good
description of $(e, e^\prime)$ data in the QE sector, reproducing
both the asymmetry of the scaling function and the transverse enhancement \cite{RGF5}.
The RGF approach is appropriate for an inclusive process where only the final
lepton is detected. On the other hand, in the NC QE scattering, only the final nucleon can be detected and the cross
section is semi-inclusive in the hadronic sector. The RGF may therefore
include channels which are not present in the experimental NC QE cross
sections, but it can also recover important contributions which are not
taken into account by other models based on the IA. For instance, in
comparison with the MiniBooNE NC QE $\nu$ and $\bar{\nu}$
scattering data~\cite{AguilarArevalo:2010cx,AguilarArevalo:2013nkf}, the RMF
generally underpredicts the experimental cross
section, while the RGF results are in reasonable agreement
with the data~\cite{Gonzalez-Jimenez:2013xpa,Meucci:2014pka}.

\subsection{Spectral function}

The SF approach is based on the factorisation {\em ansatz}, which amounts to writing the nuclear final state as a product of a plane wave, describing the motion of the
knocked out nucleon, and a $(A-1)$-nucleon state, describing the recoiling spectator system.  Within this scheme, the nuclear cross section reduces to the  convolution of the
elementary scattering cross section with the target spectral function, yielding the energy and momentum distribution of the struck nucleon in the initial state.

The spectral function of Refs.~\cite{ref:Omar_LDA,ref:Omar_oxygen}, employed in this work, accounts for the shell structure of the oxygen nucleus, extracted from experimental $(e,e'p)$ data~\cite{ref:Saclay}, as well as for the contribution of short-range correlations between nucleons, taken from theoretical calculations~\cite{ref:Omar_NM}. Those two components have been consistently combined by the authors of Refs.~\cite{ref:Omar_LDA,ref:Omar_oxygen} within the framework of the local-density approximation (LDA). The LDA approach, based on the tenet that nucleon-nucleon correlations are largely unaffected by surface and shell effects, allows one to obtain the correlation contribution for a finite nucleus from the corresponding results computed for uniform nuclear matter at different densities.

The LDA spectral functions have been successfully used in the analysis of inclusive electron-scattering data for carbon and oxygen targets at beam energies up to few GeV~\cite{ref:Omar_LDA,ref:Omar_oxygen}. Moreover, the LDA momentum distribution of carbon turns out to be  consistent with that extracted from $(e,e'p)$ data at large missing energy and missing momentum~\cite{ref:Rohe,ref:Rohe2}.

\begin{figure}
\centering
    \includegraphics[width=0.80\columnwidth]{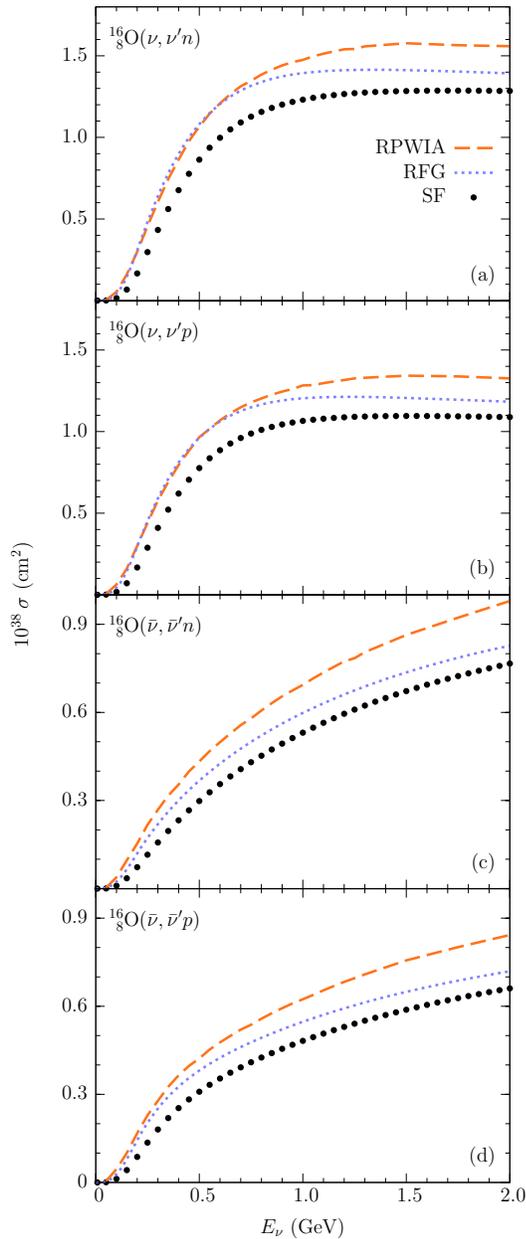}
    \subfigure{\label{fig:nu_n}}
    \subfigure{\label{fig:nu_p}}
    \subfigure{\label{fig:anu_n}}
    \subfigure{\label{fig:anu_p}}
\caption{\label{fig:CS}(Color online) Cross sections for neutron [(a) and (c)] and proton [(b) and (d)] knockout from oxygen induced by NC QE interaction of neutrino [(a) and (b)] and antineutrino [(c) and (d)]. The RPWIA and SF results are shown by the dashed lines and the filled circles, respectively. The RFG calculations, represented by dotted lines, are also included, for reference.
}
\end{figure}
\begin{figure}
\centering
    \includegraphics[width=0.80\columnwidth]{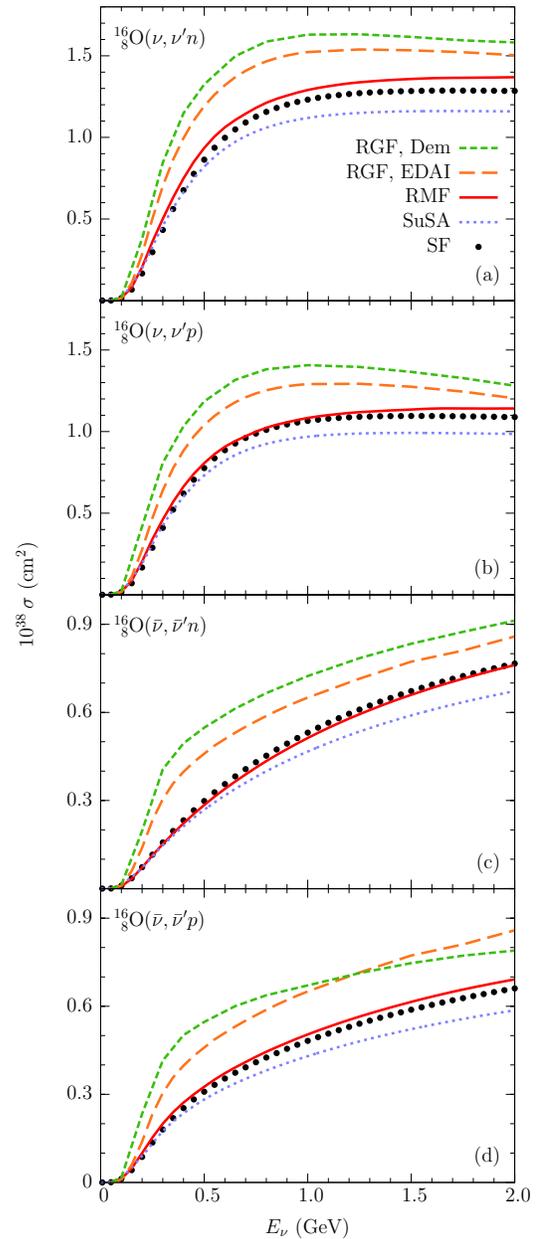}
    \subfigure{\label{fig:nu_n_FSI}}
    \subfigure{\label{fig:nu_p_FSI}}
    \subfigure{\label{fig:anu_n_FSI}}
    \subfigure{\label{fig:anu_p_FSI}}
\caption{\label{fig:CS_FSI}(Color online) Same as in Fig.~\ref{fig:CS} but for the SuSA model (dotted lines), the RMF approach (solid lines), and the RGF calculations with the EDAI (long-dashed lines) and ``democratic'' (short-dashed lines) potentials.
For comparison, we also include the SF results displayed in Fig. \ref{fig:CS}, represented by filled circles.
}
\end{figure}

\section{Results}
\label{results}

All calculations presented in this article have been  performed using the same values of physical constants, given in Refs.~\cite{ref:gamma,ref:nKnockout}. We employ the electromagnetic form factors parametrized according to Refs.~\cite{ref:Kelly,ref:Riodan}, and the dipole parametrization of the axial form factor \[F_A(Q^2)=\frac{g_A}{(1+Q^2/M_A^2)^2},\] with the coupling constant $g_A=-1.2701$~\cite{ref:PDG2012}.

For the axial mass $M_A$, we apply the value 1.03~GeV \cite{ref:Meissner}, in good agreement with that obtained in recent analyses \cite{ref:BBA03} of deuteron measurements~\cite{ref:BNL81,ref:ANL82,ref:BNL90}. This choice appears to be best suited for the purpose of the present analysis, since it allows for a clear separation between the dynamics of the elementary interaction vertex and nuclear dynamics, determining the reaction mechanism.

Note that in the studies of Ref.~\cite{ref:gamma,ref:nKnockout}, aimed at providing an estimate of \emph{measurable} cross sections including the contribution of multinucleon processes, the axial mass was instead set to the value $M_A=1.2$~GeV, determined by the K2K Collaboration~\cite{ref:Gran} for the oxygen target.

The RGF calculations have been carried out using  two
parametrizations of the relativistic optical potential of $^{16}$O:
the energy-dependent and $A$-independent
(EDAI) model  of Ref.  \cite{OP1} and  the more recent ``democratic''  phenomenological optical potential of Ref.~\cite{OP2}.
The former is a single-nucleus parametrization, constructed
to reproduce elastic proton-oxygen data, while the latter
has been obtained from a global fit to more than two hundred data sets
of elastic proton-nucleus data for a broad range of targets, from helium to lead.

The $\isotope[16][8]{O}(\nu,\nu N)$ and $\isotope[16][8]{O}(\bar\nu,\bar\nu N)$ cross sections computed using the theoretical approaches discussed in Sec. \ref{approaches} are displayed in Figs. \ref{fig:CS} and \ref{fig:CS_FSI}.

In Fig. \ref{fig:CS}, we show the cross sections obtained from the RPWIA, RFG, and SF models. Because these approaches do not include FSI,  comparing their results
provides information on the uncertainty associated with the description of the initial state.  However, it should be kept in mind that---to the extent to which the sum over the
undetected hadronic final states spans a complete set---the total inclusive cross section in the QE sector is unaffected by the inclusion of FSI.  In the SF approach this property is
fulfilled by construction.  However, in other approaches the inclusion of FSI is found to have non-negligible effects (see, e.g, \cite{TEN3,xx,yy}).

For neutrino scattering, the discrepancy between the predictions of the different approaches
is 21\% (23\%) at 0.6 (1.5) GeV,
and turns out to be similar for neutron and proton knockout.
For antineutrino scattering, the spread is more
significant,
reaching 40\% (29\%) at 0.6 (1.5) GeV for neutron knockout.
This feature is not surprising, since in this case destructive interference between the response functions makes the cross section more sensitive to differences in modeling.

In Fig. \ref{fig:CS_FSI}, we compare the results of the phenomenological  SuSA approach to those obtained from the RMF and RGF models, in which
the effects of FSI are explicitly taken into account. In the RMF model, real scalar and vector relativistic potentials are used, whereas the EDAI and ``democratic'' complex
optical potentials are employed in the RFG approach. Note that, unlike the previous ones,  the implementation of the SuSA model employed to obtain the results of Fig. \ref{fig:CS_FSI}
includes the effects of Pauli blocking, following the procedure described in Ref.~\cite{Megias:2014kia}. For comparison, we also show the results of the SF
approach, which are identical to those displayed in Fig. \ref{fig:CS}.

It is apparent that, compared to the RMF model, the SuSA approach yields
sizeably lower cross sections. This is likely to be ascribed to the fact that the
RMF predicts an enhancement of the transverse response, as an effect of off-shell
spinor distortion~\cite{Ivanov:2013bta}. The transverse enhancement, clearly observed
in electron scattering data, is not reproduced by the present version of the SuSA model, in which the
same scaling function is used in the longitudinal and transverse channels.
Work aimed at improving the SuSA approach to implement this feature is discussed in Ref.~\cite{ref:SuSAv2}.
It is interesting to observe that the RMF and SF approaches, while being based on very different
models of nuclear dynamics, yield remarkably similar results.
The two curves corresponding to the cross sections obtained from the RGF approach lay significantly above
the ones corresponding to the RMF, SuSA and SF models. In addition, they show a sizable sensitivity to the
the optical potentials, the discrepancy between the results obtained using the EDAI and ``democratic'' parametrizations ranging between $\sim$5--10\% at $E_\nu \sim 2$ GeV and
$\sim$10--25\% in the energy region $\sim$0.3--0.5 GeV.

As pointed out in Ref.~\cite{Gonzalez-Jimenez:2013xpa}, the larger cross sections in the RGF may be associated with
the redistribution of  the strength arising from reaction mechanisms other than single nucleon knockout, such as rescattering of the outgoing nucleon---possibly leading to the
excitation of non-nucleonic degrees of freedom---or scattering off a nucleon belonging to a correlated pair. These channels, although not explicitly included
in the RGF, may be phenomenologically taken into account by the imaginary part of the optical potential.

\section{Conclusions}
\label{conclusions}

We have calculated the $\isotope[16][8]{O}(\nu,\nu N)$ and $\isotope[16][8]{O}(\bar\nu,\bar\nu N)$ cross sections in the QE channel, using a variety of theoretical approaches extensively validated through comparisons to electron scattering data.

We emphasize that our analysis is not meant to assess the validity of the assumptions underlying the different approaches. The aim of this article is to quantify the uncertainty associated with the theoretical description of the nuclear cross sections,  comparing the predictions of the models outlined in Sec.~\ref{approaches}.

The spread of the theoretical results depends appreciably on energy, and turns out to be larger for antineutrino cross sections. While for the $\isotope[16][8]{O}(\bar\nu,\bar\nu p)$ knockout and $E_\nu \gtrsim 1.23$ GeV, the highest cross section is predicted by the RGF model with the EDAI potential, for all other processes and kinematics, the highest results are obtained using the RGF model with the ``democratic'' potential. In all the cases, the SuSA calculations yield the lowest cross sections.
For example, at $E_\nu = 600$ MeV, the SuSA $(\nu,\nu^\prime N)$  and  $(\bar\nu,\bar\nu^\prime N)$ results are lower than the RGF ones by $\sim$37\% and $\sim$47\%, respectively. When the energy increases, the differences are somewhat reduced, and at $E_\nu = 1.5$ GeV, the SuSA cross sections are $\sim$30--33\% lower than the RGF ones.

It is apparent that the broad spread of the results is to be mainly ascribed to the large cross sections obtained from the RGF approach. The discrepancies between the predictions of the SF, RMF, and SuSA approaches turn out to be much less pronounced, not exceeding $\sim$15\% over the energy range $0.3\leq E_\nu\leq 2.0$ GeV.

Note that the calculations discussed in this article have been performed using the dipole parametrization of the axial form factor. While another $Q^2$ dependence~\cite{ref:BBBA07,ref:nondipoleFA,ref:Megias_nondipoleFA} may introduce noneligible effects on the cross sections, an estimate of the corresponding uncertainties would require new experimental constraints, preferably from neutrino-deuteron and antineutrino-proton scattering.

We recall that uncertainty of the strange axial coupling constant, $\Delta s= -0.08\pm0.05$, translates into uncertainties of the $\isotope[16][8]{O}(\nu,\nu N)$ and $\isotope[16][8]{O}(\bar\nu,\bar\nu N)$ cross sections, shown in Ref.~\cite{ref:nKnockout} not to exceed $\sim$6\% for neutrinos and $\sim$8\% for antineutrinos. On the other hand, its effect on the $\isotope[16][8]{O}(\nu,\nu)$ and $\isotope[16][8]{O}(\bar\nu,\bar\nu)$ cross sections is less pronounced, because the proton and neutron contributions largely cancel~\cite{Gonzalez-Jimenez:2013xpa,ref:nKnockout,veneziano,Ivanov:2015wpa}.

As a final cautionary remark, we point out that multi{\-}nucleon emission processes, not taken explicitly into account in any of the approaches outlined in Sec. \ref{approaches},
are known to provide a non-negligible contribution to the nuclear cross sections in the QE channel. Their inclusion may somewhat alter the pattern emerging from our analysis.

\begin{acknowledgments}
The work of AMA has been supported by the National Science Foundation under Grant No. PHY-1352106. The work of MBB, OB, CG, and AM was supported by INFN under grant MANYBODY.
RGJ acknowledges financial help from VPPI-US (Universidad de Sevilla) and from the Interuniversity Attraction Poles Programme initiated by the Belgian Science Policy Office. GDM acknowledges support from a fellowship from the Junta de Andaluc{\'i}a (FQM-7632, Proyectos de Excelencia 2011).
\end{acknowledgments}

\end{document}